COMMUNICATION

# From Molecular Design to Optical Anisotropy: Orientation Control in BODIPY Langmuir-Blodgett Films

Lilia Huynh,[a] Jason Bessonnet,[b] Lucas Frédéric,[c] Céline Fiorini-Debuisschert,[a] Simon Vassant,[a] Hélène Fensterbank,[b] Emmanuel Allard,[b] David Kreher,*[b] Fabrice Charra,[a] Nicolas Fabre *[a]

[a] L. Huynh, Dr. C. Fiorini-debuisschert, Dr. S. Vassant, Dr. F. Charra, Dr. N. Fabre
Service de Physique de l'État condensé (SPEC)
Université Paris-Saclay, CEA-CNRS
F-91191, Gif-sur-Yvette, France
E-mail: nicolas.fabre@cea.fr

[b] J. Bessonnet, Dr. H. Fensterbank, Dr. E. Allard, Prof. D. Kreher
Institut Lavoisier de Versailles (ILV)
Université Paris-Saclay, UVSQ, CNRS
78035 Versailles, France
E-mail : david.kreher@uvsq.fr

[c] Dr. L. Frédéric
Laboratoire de Photophysique et Photochimie Supramoléculaire et Macromoléculaire (PPSM))
Université Paris-Saclay, ENS Paris-Saclay, CNRS
91190, Gif-Sur-Yvette, France

Supporting information for this article is given via a link at the end of the document.

**Abstract:** Molecular orientation within ultrathin films is a critical factor in controlling their optical and electronic properties for surface-based photonics and optoelectronics. In this study, we examine two amphiphilic boron-dipyrromethene (BODIPY) derivatives, distinguished by the number of hydrophobic alkyl chains, and investigate their organization and optical response using Langmuir–Blodgett deposition. The spatial and orientational distribution of molecules at the nanoscale is determined by combining hyperspectral imaging, photoluminescence radiation pattern analysis, and incidence-angle-resolved absorption spectroscopy. We evidence, both experimentally and based on a new and original theoretical model, the formation of organized monolayers exhibiting either in-plane or perpendicular transition dipole alignment, depending on the molecular symmetry. These results underscore the deep impact of molecular engineering on supramolecular order and anisotropic optical properties at interfaces, providing a robust strategy for the design of functional thin films down to the monolayer level for advanced optical devices.

The control of molecular orientation within thin films represents a key challenge in the design of advanced functional materials, given its significant influence on their electronic, mechanical, and optical properties.[1–3] The relative orientation of molecules strongly affects optical behavior, as exemplified by cyanine dyes, where side-by-side (H-type) and head-to-tail (J-type) arrangements lead to distinct excitonic couplings, either quenching or enhancing radiative recombination.[4–7] Furthermore, emission can be modulated through phenomena such as aggregation-caused quenching (ACQ) or aggregation-induced emission (AIE).[8,9] Interestingly, anisotropic molecular arrangements have the capacity to induce anisotropic responses, which are critical for a variety of applications, including organic light-emitting diodes, photovoltaic devices, sensors, metamaterials and nonlinear optics.[10–14] A multitude of techniques have been reported in the literature as methods for fabricating dense molecular films. These include drop-casting, thermal-sublimation, and melt processing as a few examples.[15–17] Among these, Langmuir–Blodgett (LB) deposition is distinguished by its ability to transfer well-defined monolayers formed at the air–water interface onto solid substrates.[18,19] This method enables precise control over molecular packing density and orientation, which can be tuned through molecular design and processing conditions. In particular, side edge-on or vertical head-on orientations – which are relevant for electronic and photonic applications – appear to be attainable.[3,20,21] In this context, optical anisotropy in Langmuir-Blodgett films has been extensively studied using reflection-based techniques such as Reflectance Anisotropy Spectroscopy (RAS)[22–26] and Differential Reflection Spectroscopy (DRS).[27] While these methods are highly sensitive to the complex dielectric function, they often require sophisticated analysis to decouple dispersive and absorptive contributions. In the realm of organic dyes, boron-dipyrromethene (BODIPY) derivatives have emerged as promising candidates towards optoelectronic applications due to their high fluorescence quantum yields, excellent photostability, and facile functionalization for tuning their optical properties.[28–31] However, the relationship between subtle structural modifications and the resulting supramolecular organization and orientation in LB films remains poorly understood and characterized, despite its critical influence on optical anisotropy and device performance.

In order to bridge this gap, we designed and synthesized two BODIPY derivatives (see Scheme 1), sharing the same chromophore core but differing in their molecular symmetry and amphiphilic balance through the number and distribution of long alkyl chains. The BODIPY asymmetric derivative $\mathbf{B_A}$, bearing a single alkyl chain, exhibits a more pronounced amphiphilic character, whereas the symmetric BODIPY compound $\mathbf{B_S}$, with two equivalent chains, presents a more balanced interfacial affinity. In order to accurately determine their molecular orientation at the surface, we employed advanced optical characterization techniques, and we extended a previous analytical model.[21] In the case reported here, the use of transparent substrate enables angle-resolved transmission spectroscopy, which allows for a direct probe of the absorption component, thereby simplifying the quantitative determination of molecular tilt angles. This “optical-only” approach is proposed here as a robust and accessible alternative to X-ray scattering methods, which, although standard, require complex facilities and specialized data treatment. The present study details the preparation, characterization and orientation analysis of monolayers composed of these new BODIPY derivatives, both on an experimental and theoretical point of view. We demonstrate in particular that the molecular structure, symmetry and amphiphilic character have a drastic influence on the molecular orientation leading to marked differences in the thin film photonic properties.

**Scheme 1.** Chemical structures and synthetic routes of BODIPY $B_A$ and $B_S$. (a) 1) THF, n-BuLi (2.5 M in hexane, 1 eq), -78 °C 2) 1-iodododecyl (1 eq), -78 °C (b) 1) THF, n-BuLi (2.5 M in hexane, 1.1 eq), -40 °C 2) B(OMe)3 (3 eq), -90 °C 3) Pinacole (5 eq) RT (c) DCM, DMF, NIS (1.1 eq), 0 °C to RT (d) THF:H2O, Pd(PPh3)4 (0.1 eq), Cs2CO3 (5 eq), (2) (1.6 eq), 85 °C (e) DCM, NIS (4 eq), RT (f) THF:H2O, Pd(PPh3)4 (0.2 eq), Cs2CO3 (10 eq), (2) (2.4 eq), 85 °C

The structures of the amphiphilic BODIPY derivatives $\mathbf{B_A}$ and $\mathbf{B_S}$ and the synthetic routes are depicted in Scheme 1. Both were obtained via Suzuki–Miyaura cross-coupling between halogenated BODIPY intermediates **3** and **4** respectively and thiophene boronate esters **2** bearing a dodecyl chain, after optimizing the reaction conditions (catalyst and its ratio, solvent, temperature). The use of mono *versus* bi-substituted BODIPY precursors affords asymmetric ($\mathbf{B_A}$) and symmetric ($\mathbf{B_S}$) molecular targets, respectively, which were fully characterized in terms of chemical structures (see Supporting Information, part 1.2, Figures S1 to S8). To sum up, this specific and quite straightforward molecular design allows a direct comparison between a symmetric and an asymmetric BODIPY architecture, the latter exhibiting a stronger amphiphilic character due to the unbalanced distribution of hydrophobic and hydrophilic moieties. Such a difference in molecular symmetry is expected to play a decisive role in dictating interfacial orientation, promoting an upright arrangement for the asymmetric derivative and a planar organization for the symmetric one.

Thus, we prepared monolayers of each compound using the Langmuir–Blodgett technique. The compression isotherms for $\mathbf{B_A}$ and $\mathbf{B_S}$ are presented in Figure S9. A comparison of the general shapes of the isotherms reveals notable similarities. The initial flat region at high molecular area corresponds to free molecules lying at the air–water interface, while the sharp rise observed below 60 and 50 Å² molecule⁻¹ for $\mathbf{B_A}$ and $\mathbf{B_S}$, respectively, indicates the onset of the condensed 2D phase. The monolayers were transferred onto glass substrates at a surface pressure of 20 mN·cm⁻¹, corresponding to areas of 30 Å² molecule⁻¹ for $\mathbf{B_A}$ and 35 Å² molecule⁻¹ for $\mathbf{B_S}$, yielding a constant transfer ratio of 1. Regarding the durability of the prepared LB films, it is worth noting that while the BODIPY derivatives exhibit intrinsic chemical robustness, the stability of a single monolayer remains sensitive to environmental factors. Preliminary observations suggest that photostability under high-intensity illumination is a limiting factor, this is why the spectroscopic measurements presented here were performed at low excitation power, with integration times kept to a minimum.

Then, the absorption and emission spectra of the two compounds in dichloromethane solutions were recorded (black curves in Figure 1). In addition, we recorded hyperspectral absorption images in normal incidence along a 230 µm line for $\mathbf{B_A}$ and $\mathbf{B_S}$, as shown in Figure S10, with the corresponding averaged spectra displayed (orange and red curves respectively in Figure 1). The images obtained reveal distinct spatial characteristics depending on the compound.

First of all, on a general point of view and in comparison, with its spectrum in solution, the monolayer of $\mathbf{B_A}$ demonstrates a hypsochromic shift in both absorption and emission (Figure 1a), which is indicative of the formation of H-aggregate-like dye organizations.[7] On the contrary, still comparing with the solution but in the case of $\mathbf{B_S}$, both the absorption and emission spectra manifest bathochromic shifts (Figure 1b), thereby indicating J-aggregate-like dye organization within the $\mathbf{B_S}$ monolayer.[7]

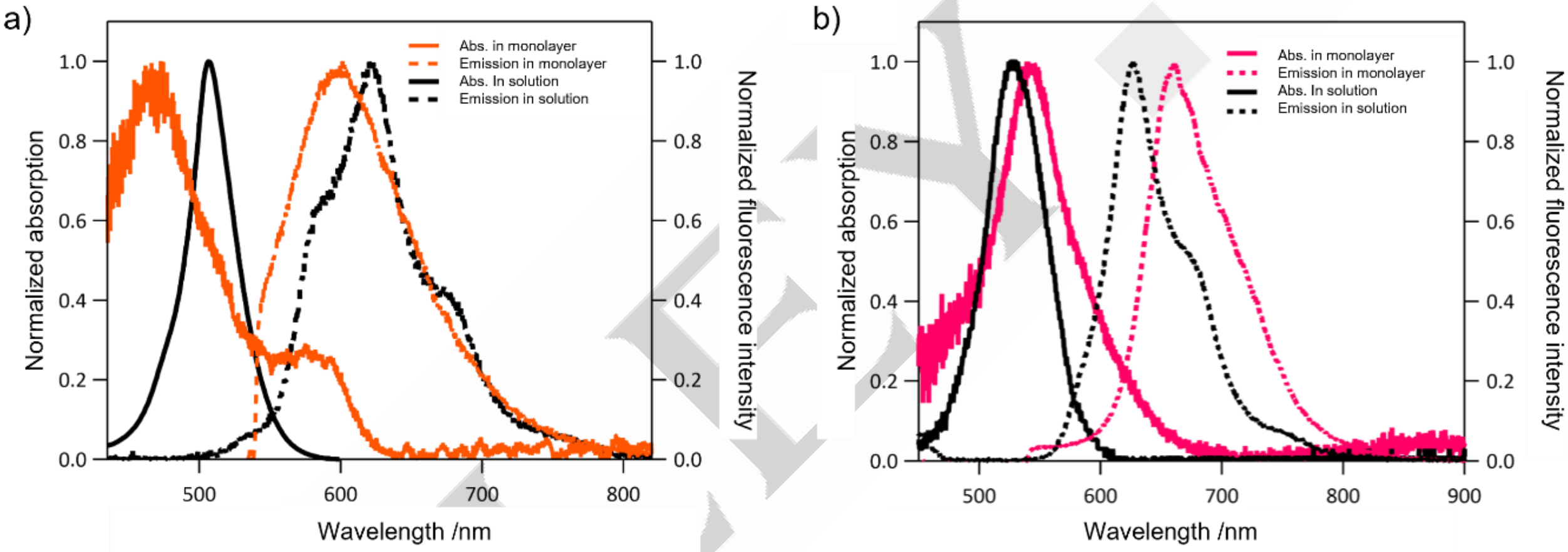

**Figure 1.** Normalized absorption (full lines) and emission (dashed lines) spectra in solution (black lines, $1\times10^{-5}$ mol L$^{-1}$ in dichloromethane, optical path 0.2 cm) and in the monolayer resulting from the LB deposition (colored lines) for a) $B_A$ (orange lines) and b) $B_S$ (purple lines).

Then, going into details, in the case of $\mathbf{B_A}$ only a very weak absorption band is visible over the entire measured line. After averaging, as shown in Figure 1a, this weak contribution consolidates into a broad band peaking at 465 nm on the one hand, and on the other hand with the presence of an additional band of weaker intensity presenting a maximum wavelength at 580 nm and which could be characteristic of a charge transfer band due to molecular arrangement. At 465 nm wavelength, the monolayer absorbs 3% of the incident light. Utilizing the molar absorption coefficient, from molecules in $CH_2Cl_2$ solution, of $\varepsilon_{(\mathbf{B_A})}$~78×10³ mol⁻¹ L cm⁻¹, yield the absorption cross-section $\sigma_{(\mathbf{B_A})}$~2.9×10⁻¹⁶ cm² molecule⁻¹, for isotropic orientation of the molecules. An apparent area of 95 Å² per molecule is further derived from the absorbance and cross-section. This value is significantly higher than the area defined during LB deposition. This discrepancy indicates a molecular organization leading to suboptimal light absorption at normal incidence. The following sections will provide further support for this hypothesis. Moreover, the $\mathbf{B_A}$ monolayer exhibits fluorescence ranging from 530 to 800 nm, with a maximum at 600 nm under 532 nm excitation

Unlike $\mathbf{B_A}$, the hyperspectral image of $\mathbf{B_S}$ displays a homogeneous absorption band across the probed area, with no detectable defects down to the optical resolution (~0.5 µm). The corresponding averaged spectrum is shown in Figure 1b. The monolayer absorbs within the 470-680 nm range, with a maximum peaking at 540 nm. At this particular wavelength, the film absorbs approximately 10% of the incident light, a three-fold increase as compared to $\mathbf{B_A}$. We estimate the cross-section at $\sigma_{(\mathbf{B_S})}$~2.0×10⁻¹⁶ cm² molecule⁻¹, based on a molar absorption coefficient of $\varepsilon_{(\mathbf{B_S})}$~55×10³ mol⁻¹ L cm⁻¹. From these values, we calculate an area of 25 Å² molecule⁻¹, which aligns well with the area defined during LB deposition and confirms efficient molecular packing and orientation for normal incidence absorption. As with the $\mathbf{B_A}$ monolayer, the $\mathbf{B_S}$ monolayer exhibits fluorescent properties as well, with an emission spectrum ranging from 590 to 800 nm and peaking at 660 nm for a 532 nm excitation.

To complement the optical homogeneity observed via hyperspectral imaging, Atomic Force Microscopy (AFM) measurements were performed. The topographic maps, displayed in Figure S13, show a surface where the roughness is dictated by the glass substrate. The absence of detectable aggregates at this scale supports the formation of a continuous and uniform molecular layer.

Based on these preliminary results and in order to ascertain the orientation of the molecules on the surface, we employed two complementary techniques: photoluminescence radiation pattern analysis and incidence-angle-resolved absorption spectroscopy. As further shown below, the experimental observations were systematically compared to simulations corresponding to either in place or out of plane molecular arrangements.

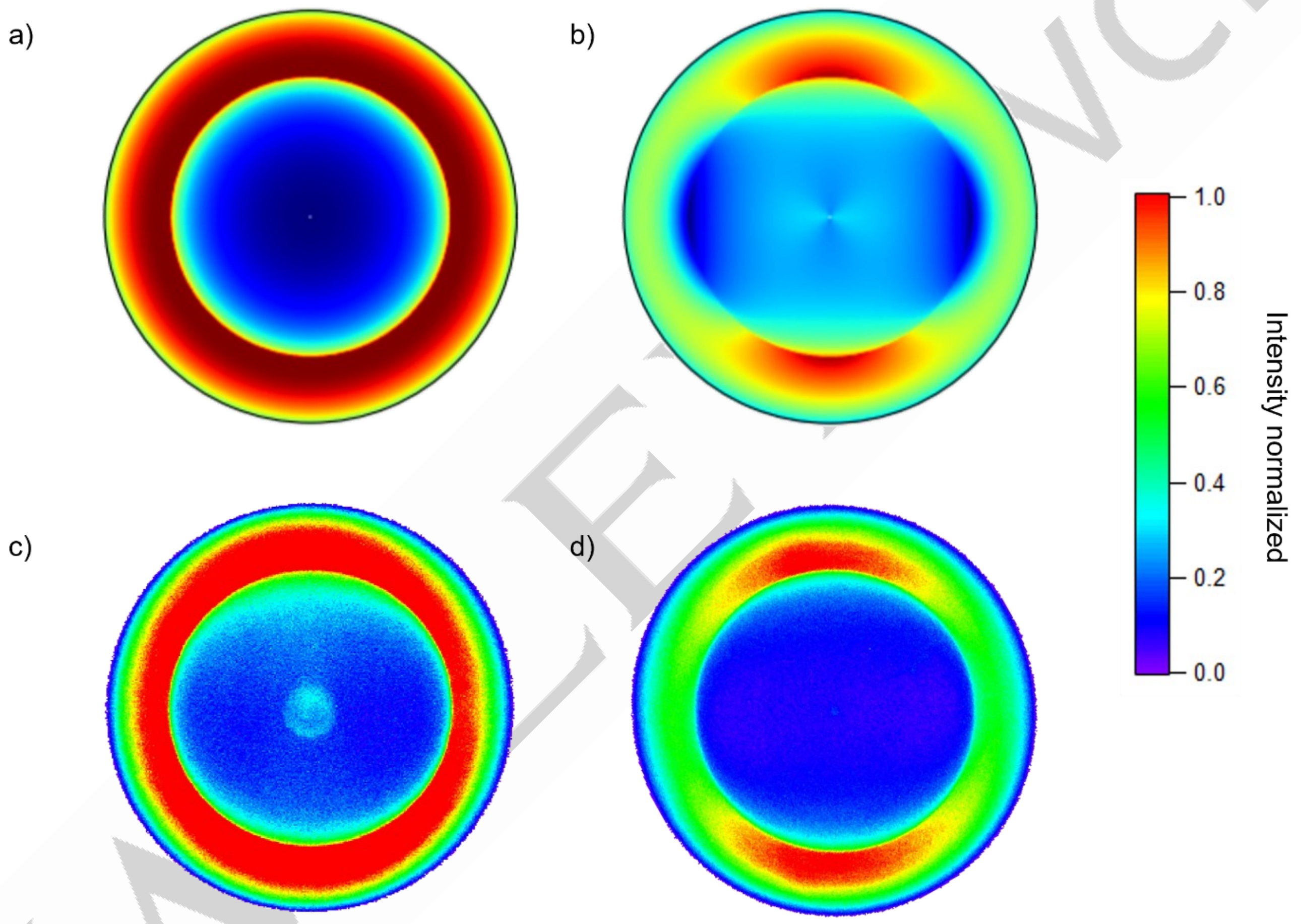

**Figure 2.** Calculated emission patterns of a single dipole at an air-glass interface in the back focal plane of a N.A.=1.5 objective lens for a) a dipole perpendicular to the plane and b) an in-plane dipole. Observed pattern for c) $\mathbf{B_A}$ and d) $\mathbf{B_S}$ monolayers under excitation at 532 nm for p-polarization.

The photoluminescence radiation pattern, collected in the back focal plane of a high numerical aperture microscope objective, provides insight into the orientation of the transition dipole moments of the molecules.[32] For BODIPY derivatives, both the $S_0 \rightarrow S_1$ absorption and $S_1 \rightarrow S_0$ emission transition dipole moments are collinear with the N–N axis (Figure S11). The simulated radiation patterns for a single dipolar emitter oriented either perpendicular (out-of-plane) or parallel (in-plane) to the substrate are shown in Figures 2a and 2b, respectively.[32] In the case of perpendicularly oriented dipoles, it gives toroidal patterns with no emission detected at the center (i.e. normal emergence). Meanwhile, in-plane dipoles yield two intensity lobes with minima at 90 °.

Within the experimental framework, the samples were excited at an incidence angle of 45°, to excite both in-plane and out-of plane molecules. The illuminated area is large, filling the full objective field of view. We present the experimental radiation patterns collected under *p*-polarized excitation for $\mathbf{B_A}$ and $\mathbf{B_S}$ in Figures 2c and 2d, respectively. In the case of *p*-polarization, both in-plane and out-of-plane dipoles are excited with nearly equal probability (~45% and ~55%, respectively). The radiation pattern exhibited by the $\mathbf{B_A}$ monolayer is analogous to the simulated pattern for out-of-plane dipoles, while that of $\mathbf{B_S}$ closely corresponds to the in-plane configuration. The results obtained from this study indicate that the $\mathbf{B_A}$ molecules are primarily oriented perpendicular to the substrate, while the $\mathbf{B_S}$ molecules are predominantly aligned parallel to the surface. In order to provide further support for this conclusion, line profiles along the principal axes of the experimental radiation patterns were compared to theoretical profiles (see Figure S12). For both monolayers, the positions of local maxima are consistent with theory, indicating a satisfactory match. However, the minima in the experimental profiles are less pronounced than in the theoretical case, particularly in the direction orthogonal to the transition dipole moment. The observed discrepancies are likely attributable to the distribution of azimuthal orientations within the monolayer, resulting in incomplete extinction along the anticipated nodal directions. Notwithstanding this observation, the prevailing pattern substantiates the hypothesis that the transition dipoles in $\mathbf{B_S}$ are aligned parallel to the glass substrate, while those in $\mathbf{B_A}$ are predominantly perpendicular.

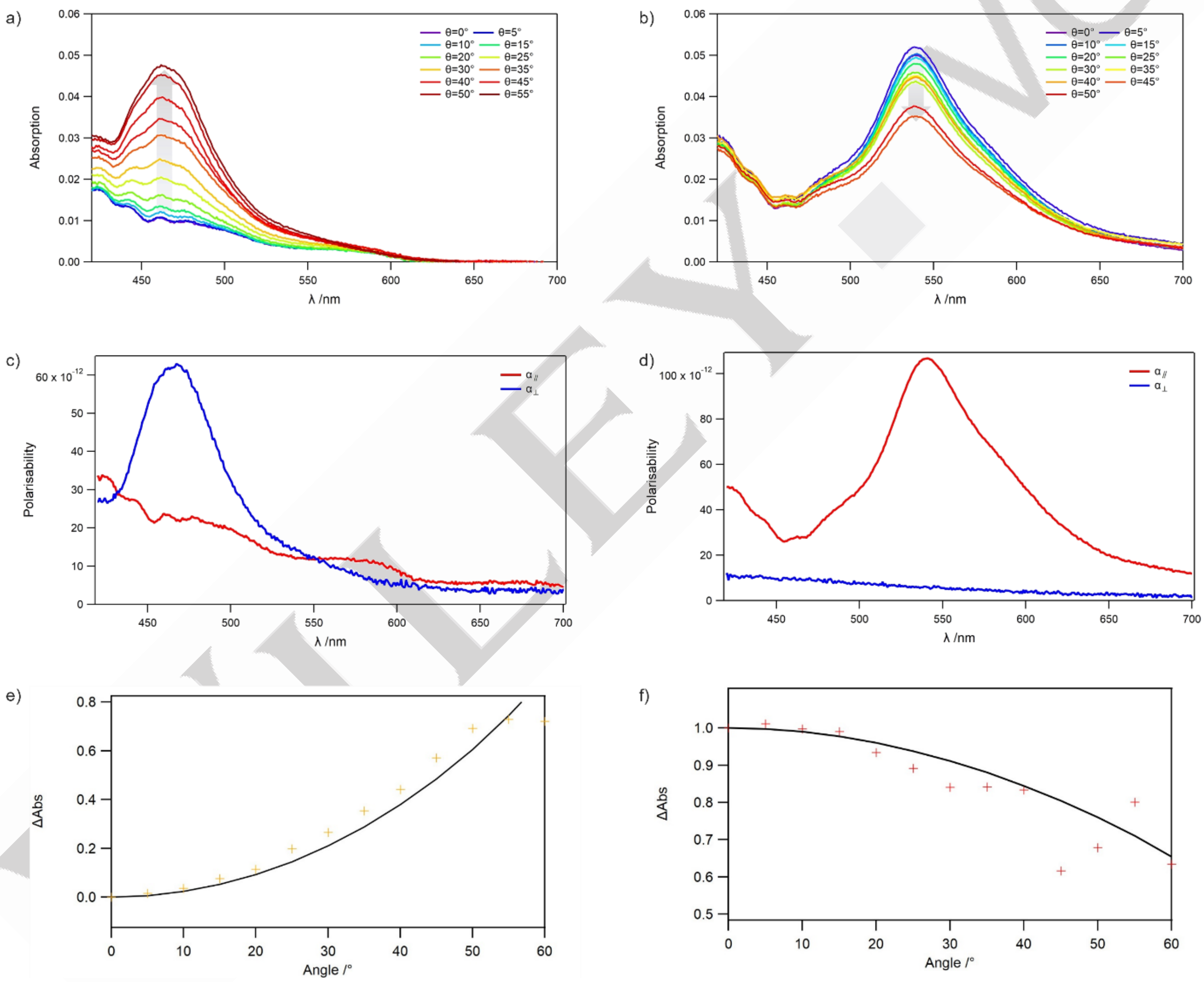

**Figure 3.** Absorption spectra of a) $\mathbf{B_A}$ monolayer and b) $\mathbf{B_S}$ monolayer at several incidence angle. Polarizability contribution in the plane ($\alpha_{\parallel}$, red line) and perpendicular to the plane ($\alpha_{\perp}$, blue line) of c) $\mathbf{B_A}$ monolayer and d) $\mathbf{B_S}$ monolayer. e) Evolution of the absorption of $\mathbf{B_A}$ monolayer at 480 nm (orange cross) as function of incidence angle with its theoretical evolution (black line). f) Evolution of the absorption of $\mathbf{B_S}$ monolayer at 540 nm (purple cross) as function of incidence angle with its theoretical evolution (black line).

In order to corroborate these findings, we conducted incidence-angle-resolved absorption spectroscopy. We used a linearly polarized (*p*-polarized, TM) white-light beam to probe the monolayer-coated glass substrates. We record absorption spectra at various angles of incidence, θ. Figure 3a and 3b shows the spectra for $\mathbf{B_A}$ and $\mathbf{B_S}$ respectively.

At normal incidence (θ=0 °), only in-plane dipoles absorb light. As the angle of incidence increases, the electric field component perpendicular to the substrate also increases, progressively allowing excitation of out-of-plane dipoles.

For $\mathbf{B_A}$, we observe a negligible absorption at normal incidence, confirming the poor in-plane alignment as previously hypothesized from the microspectroscopy analyses (Figure 1). However, as the parameter θ increases, a pronounced absorption band manifests, similar to the micro-absorption spectrum. This behavior confirms that the absorption is predominantly attributable to transition dipoles oriented perpendicularly to the substrate.

To interpret these spectra, we admit that the response of the molecular monolayer is local, uniform and surfacic. This is legitimate since its thickness and intermolecular distances are much smaller than the wavelength. Given the geometry of the system, the optical response of the monolayer can be represented by the in-plane and out-of-plane components of the surface polarizability tensor, $\alpha_\parallel$ and $\alpha_\perp$.[21] Their imaginary parts $Im(\alpha_\parallel)$ and $Im(\alpha_\perp)$ can then be extracted from incidence-dependent absorption measurements using the following equations:

From these spectra, the in-plane and out-of-plane components of the imaginary part of the surface polarizability tensor, $Im(\alpha_\parallel)$ and $Im(\alpha_\perp)$, were extracted using the following equations:[21]

$$Im(\alpha_\parallel) = \frac{\lambda\varepsilon_0}{2\pi}\mathrm{Abs}(\lambda, \theta = 0)(n_1 + n_2) \qquad \text{(Eq.1)}$$

$$Im(\alpha_\perp) = \left[\frac{\lambda\varepsilon_0}{2\pi}\mathrm{Abs}(\lambda, \theta = 0) - \frac{Im(\alpha_\parallel)}{\frac{n_1}{\cos\theta} + \frac{n_2}{\cos\varphi}}\right] \times \frac{\frac{\cos\theta}{n_1} + \frac{\cos\varphi}{n_2}}{n_1 \sin\theta\, n_2 \sin\varphi} \qquad \text{(Eq.2)}$$

where λ is the wavelength, $\varepsilon_0$ is the vacuum permittivity, Abs(λ, θ) is the measured absorbance, and $n_1$ and $n_2$ are the refractive indices of air and the substrate, respectively. φ is related to θ by Snell's law. The resulting $Im(\alpha_\parallel)$ and $Im(\alpha_\perp)$ for $\mathbf{B_A}$ are shown in Figure 3c. The result of this calculation is that $Im(\alpha_\parallel)$ is negligible, indicating insignificant in-plane absorption. Conversely, $Im(\alpha_\perp)$ reproduces the absorption spectrum, thereby confirming that the out-of-plane component solely contributes to absorption and, consequently, that the $\mathbf{B_A}$ dipoles are predominantly perpendicular to the surface. In order to provide further validation of this result, we introduce the variation of absorption ΔAbs(θ) induced by $Im(\alpha_\perp)$ and $Im(\alpha_\parallel)$ as a function of incidence angle θ relative to that at θ =0. ΔAbs is practically defined in detailed in Supporting Information section 2.5. Normalized ΔAbs(θ) at 465 nm was plotted as a function of θ (Figure 3e). The absorption increases with θ, in agreement with the behavior expected for out-of-plane dipoles. We fit this evolution using a new and original simplified form of the angle-dependent absorption expression, introduced in this work:

$$\Delta\mathrm{Abs}(\theta) = \frac{n_1 \sin\theta\, n_2 \sin\varphi}{\frac{\cos\theta}{n_1} + \frac{\cos\varphi}{n_2}} Im(\alpha_\perp) + \frac{n_1 + n_2}{\frac{n_1}{\cos\theta} + \frac{n_2}{\cos\varphi}} Im(\alpha_\parallel) \qquad \text{(Eq.3)}$$

Assuming $Im(\alpha_\parallel) = 0$ for $\mathbf{B_A}$, this simplifies to:

$$\Delta\mathrm{Abs}(\theta) = \frac{n_1 \sin\theta\, n_2 \sin\varphi}{\frac{\cos\theta}{n_1} + \frac{\cos\varphi}{n_2}} Im(\alpha_\perp) \qquad \text{(Eq.4)}$$

The experimental angle-dependent absorption in Figure 3e is very well fitted by considering an out-of-plane only response (that is $Im(\alpha_\parallel) = 0$). Actually, the corresponding theoretical curve related to (Eq.4), displayed as a solid black line in Figure 3e, exhibits a high degree of compatibility with the data. The out-of-plane only contribution to absorption can be explained by considering a distribution of independent $\mathbf{B_A}$ molecules with transition dipole moments having an average orientation perpendicular to the substrate. Although local field effects and intermolecular couplings (e.g., H- or J-aggregation) become dominant in determining absorption spectra, the independent molecule approximation retains most characteristics of optical anisotropy, the transition dipole moments of collective excitations being a linear combination of those of its constituents. It confirms the perpendicular orientation for $\mathbf{B_A}$. and further substantiates a perpendicular molecular orientation for $\mathbf{B_A}$.

The same analysis was performed for $\mathbf{B_S}$ (Figure 3b). At normal incidence, strong absorption is observed at 540 nm, similar to the absorption spectrum obtained in micro-spectroscopy (Figure 1.a). Conversely, as the incidence angle increases, the absorption decreases, a phenomenon that stands in contrast to the behavior observed with $\mathbf{B_A}$. Calculated polarizability components (Figure 3d) demonstrate a non-zero $Im(\alpha_{\parallel})$ that corresponds to the structure of the monolayer absorption spectrum, whilst $Im(\alpha_{\perp})$ maintains a null value at all angles. This observation further supports the hypothesis that the observed absorption is attributable to in-plane transition dipole moments, thereby indicating that $\mathbf{B_S}$ molecules are aligned parallel to the substrate. Figure 3f presents the normalized variation of absorption ΔAbs at 540 nm as a function of θ for $\mathbf{B_S}$. The absorption decreases with increasing angle, consistent with the absence of out-of-plane contributions. This trend can be fitted using:

$$\Delta \mathrm{Abs}(\theta) = \frac{1}{\frac{n_1}{\cos\theta} + \frac{n_2}{\cos\varphi}} Im(\alpha_{\parallel}) \qquad \text{(Eq.5)}$$

The model displays a quite satisfying match good agreement with the data and definitively confirms that the transition moment of $\mathbf{B_S}$ molecules occurs in the plane of the substrate. The excellent agreement between the experimental data and the analytical model confirms the robustness of both our instrumental and theoretical approaches, establishing the distinct molecular orientations of the two derivatives without ambiguity. These outcomes demonstrate that molecular symmetry and amphiphilic balance are driving forces that control interfacial organization, with asymmetric architecture ($\mathbf{B_A}$) leading to an upright arrangement and symmetric geometry ($\mathbf{B_S}$) promoting in-plane alignment.

In summary, we performed absorption micro-spectroscopy, polarized fluorescence emission diagram measurements, and angle-resolved polarized absorption spectroscopy on LB-deposited monolayers of two BODIPY derivatives with distinct molecular symmetries and amphiphilic characters. The results permit to assign unambiguously the dye orientation within the film. The asymmetric derivative $\mathbf{B_A}$, featuring a single hydrophobic chain and a stronger amphiphilic character, forms upright monolayers with transition dipole moments oriented predominantly perpendicular to the substrate. In contrast, the symmetric compound $\mathbf{B_S}$, bearing two alkyl chains, assembles into a film with highly homogeneous molecular orientation with in-plane dipole alignment. While studies of this kind are typically limited to dipoles oriented parallel to the substrate, the present results provide, to the best of our knowledge, the first experimental demonstration of incidence-angle-resolved absorption spectroscopy applied to a molecular dipole oriented perpendicular to the interface. In addition, this work introduces and experimentally validates a new simplified theoretical model for angle-dependent absorption, enabling a quantitative analysis of molecular dipoles with arbitrary orientation with respect to the interface. These findings highlight the crucial role of molecular symmetry and amphiphilic balance in governing interfacial self-assembly. Furthermore, they represent a significant step forward in the quantitative characterization of anisotropic molecular monolayers, granting direct access to both in-plane and out-of-plane polarizability components. Overall, this work highlight how molecular design can be used to tailor supramolecular alignment and tune the optical response of functional monolayers at the nanoscale. The reliability of our methodology is supported by the clear consistency between the molecular design, the experimental spectroscopic signatures, and the theoretical model. By accurately identifying the transition from perpendicular to parallel dipole alignment without the need for complex scattering techniques, this approach provides a practical and robust toolkit for the structural characterization of functional monolayers. While this study successfully demonstrates the correlation between molecular symmetry and organization properties using two model derivatives, it opens the way for more systematic investigations towards general design rules. Future work will expand this library by varying alkyl chain lengths and incorporating specific functional groups to explore the influence of weak supramolecular interactions on film organization.

## Supporting Information

The authors have cited additional references within the Supporting Information.[27,28]

## Acknowledgements

The authors acknowledge the support of the French Agence Nationale de la Recherche (ANR), under grants ANR-21-CE06-0041 (project LESOMMETA)

**Keywords:** Langmuir-Blodgett • Monolayer • Absorption • Fluorescence

**Entry for the Table of Contents**

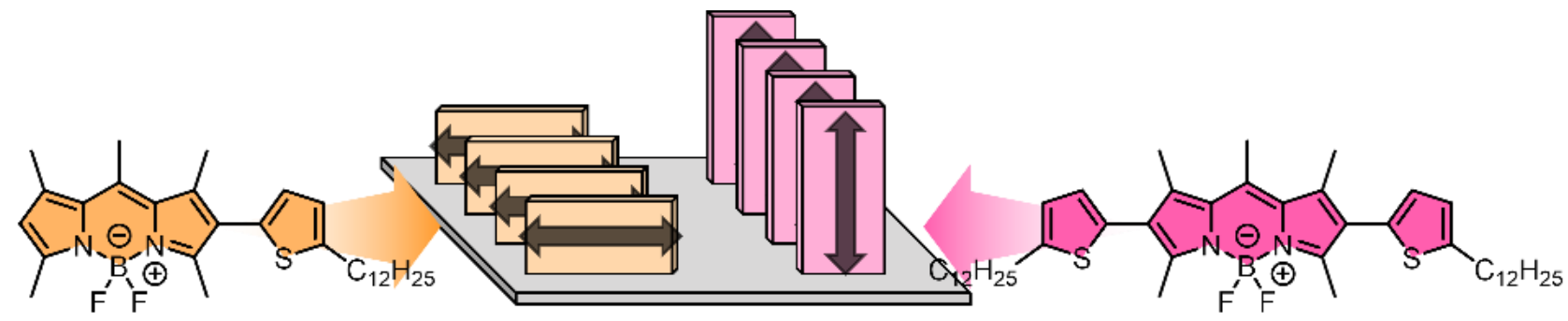

The supramolecular organization of amphiphilic boron-dipyrromethene (BODIPY) dyes in Langmuir–Blodgett monolayers is shown to govern their optical anisotropy. By correlating nanoscale imaging and angle-resolved spectroscopy, we reveal symmetry-driven control of transition dipole orientation, enabling either in-plane or out-of-plane optical responses.